\documentclass[12pt]{article}
\usepackage{amssymb,amscd,array}
\catcode `\@=11
\@addtoreset{equation}{section}

\def\qed{\blacksquare}

\newcommand{\be}{\begin{equation}}
\newcommand{\ee}{\end{equation}}
\newcommand{\bea}{\begin{eqnarray}}
\newcommand{\eea}{\end{eqnarray}}

\newcommand{\R}{\mathbb{R}}

\newcommand{\C}{\mathbb{C}}
\newtheorem{thm}{Theorem}[section]

\newtheorem{lemma}[thm]{Lemma}
\newtheorem{cor}[thm]{Corollary}
\newtheorem{prop}[thm]{Proposition}
\begin{document}
\begin{titlepage}
\begin{center}
{\bf \Large{Quantum Extended Supersymmetries\\}}
\end{center}
\vskip 1.0truecm
\centerline{Dan Radu Grigore\footnote{e-mail: grigore@theor1.theory.nipne.ro}}

\centerline{Dept. Theor. Phys., Inst. Atomic Phys.,}
\centerline{Bucharest-M\u agurele, MG 6, Rom\^ania}
\vskip 0.5cm
\centerline{G\"unter Scharf\footnote{e-mail: scharf@physik.unizh.ch}}

\centerline{Institute of Theor. Phys., University of Z\"urich}
\centerline{Winterthurerstr., 190, Z\"urich CH-8057, Switzerland}
\vskip 2cm
\bigskip \nopagebreak
\begin{abstract}
\noindent
We analyse some quantum multiplets associated with extended supersymmetries.
We study in detail the general form of the causal (anti)commutation relations.
The condition of positivity of the scalar product imposes severe restrictions
on the (quantum) model. It is problematic if one can find out quantum
extensions of the standard model with extended supersymmetries.
\end{abstract}
PACS: 11.10.-z, 11.30.Pb

\newpage\setcounter{page}1
\end{titlepage}
\section{Introduction}
The construction of a model for the interaction of elementary particles
should have the ultimate goal of providing a quantum model. Indeed, from
the phenomenological point of view a classical Yang-Mills field is
without relevance. The standard model of the elementary particles
\cite{Wei} is  basically a quantum model: to check it experimentally one
needs only the  Feynman rules i.e. the expressions for the propagators
and the vertices. But to give the propagators of some model is
equivalent to specify the set of quantum fields relevant for the model
and the expression of the vertices is nothing but the interaction
Lagrangian. So the phenomenological point of view fits very nicely with
the Bogoliubov version of perturbation theory; in this approach the
basic input is a given Hilbert space 
${\cal H}$
(which is taken of Fock type generated from the vacuum state by some set
of free quantum fields) and the interaction Lagrangian (which is some Wick
polynomial in these fields). So, all considerations of the standard model as
classical field theory followed by some quantization procedure should be viewed 
only as some auxiliary steps leading to the quantum model. It is not at all
clear if a quantization procedure really provides a Hilbert space description
of the classical model. This has to be checked in detail after the quantum
model has been constructed. In particular the positivity of the scalar product
is essential: otherwise we would have negative probabilities for some
transition processes. 

In two preceding papers \cite{GS1}, \cite{GS2} we have examined critically
the possible supersymmetric extensions of the standard model for the
case un-extended supersymmetry i.e. 
$N = 1$.
The main point was that the construction of an quantum supersymmetric
multiplet is more restrictive that the classical construction. The main
difference lies in the requirement that the multiplet has a {\it bona fid\ae}
representation in a Hilbert space. The formal construction of a quantum
supersymmetric multiplet is done by applying some (free) quantum fields on the
vacuum state: in the language of axiomatic field theory this means that we
construct the Borchers algebra (see for instance \cite{IAS}). However, one
has to provide a positively defined scalar product: only in this case
one gets from the Borchers algebra a Hilbert space by a standard procedure.
For free fields the scalar product can be obtained from the form of the
causal (anti)commutation relations using K\"allan-Lehman representation
theorem. But the supersymmetric algebra puts severe restrictions on the most
general form of these causal (anti)commutators and it is not guaranteed that
the positivity can be always enforced. In \cite{GS2} we have found out that
this imposes severe restrictions one the free parameters of the model: in
particular the vector model has a Hilbert space representation only for
positive mass. 

In this paper we investigate critically other supersymmetric models based
on extended supersymmetries. We are interested only in models which can be
used for a supersymmetric extension of the standard model (SM) and without 
particles of spin higher that $1$. From the analysis of the irreducible 
representations of the supersymmetric algebra it is known that there are only 
five multiplets describing irreducible representations with particles of spin
$s \leq 1$ namely (see for instance \cite{BSS}): for
$N = 1$
the chiral multiplet and the vector multiplet; for
$N = 2$
the hyper-multiplet and a vector multiplet; for
$N = 4$
a vector multiplet. At the level of classical field theory these models have
been studied in detail: see for instance \cite{Fa}, \cite{GSO}, \cite{HST}, 
\cite{SSW1}, \cite{SSW2}.

In this paper we analyze in the spirit of \cite{GS2} the last three cases. We 
find out that there are considerable difficulties to construct supersymmetric 
extensions of the standard model for
$N = 2$
and
$N = 4$
contrary to what it is asserted in the literature.

In Section \ref{ext} we provide a general discussion of extended supersymmetry
for quantum models. In Section \ref{n2} we consider a
$N = 2$
extended model without central charges containing spin $1$ particles. In
Section \ref{hypermultiplet} we consider the so-called hyper-multiplet which can be
used to describe matter. In Section \ref{n4} we study a
$N = 4$
extended model without central charges containing spin $1$ particles. 
In principle the results from Sections \ref{n2} and \ref{n4} could be
used for construction supersymmetric extensions of the standard model of
elementary particles \cite{Wei} but we point out some problems in achieving
this goal.
\section{Extended Supersymmetries\label{ext}}

We remind here the definition of a extended supersymmetric theory in a
pure quantum  context.   We use the notations from \cite{GS1}.

The conventions are the following: (a) we use summation over dummy indices;
(b) we raise and lower Minkowski 
indices with the Minkowski pseudo-metric
$g_{\mu\nu} = g^{\mu\nu}$
with diagonal
$1,-1,-1,-1$;
(c) we raise and lower Weyl indices with the anti-symmetric 
$SL(2,\C)$-invariant
tensor
$\epsilon_{ab} = - \epsilon^{ab}; \quad \epsilon_{12} = 1$;
(d) we denote by
$\sigma^{\mu}$
the usual Pauli matrices with elements denoted by
$\sigma^{\mu}_{a\bar{b}}$
and the convention
$\sigma^{0} = {\bf 1}$.

Suppose that we have a quantum theory of {\bf free} fields; this means that
we have the following construction:
\begin{itemize}
\item
${\cal H}$
is a Hilbert space of Fock type (associated to some one-particle Hilbert space
describing some choice of elementary particles) with the scalar product
$(\cdot, \cdot)$;
\item
$\Omega \in {\cal H}$
is a special vector called the vacuum;
\item
$U_{a,A}$ 
is a unitary irreducible representation of 
$inSL(2,\C)$
the universal covering group of the proper orthochronous Poincar\'e group such
that 
$a \in \R^{4}$ 
is translation in the Minkowski space and 
$A \in SL(2,\C)$;
\item
$U \mapsto V_{U}$
is a unitary representation of the compact group $G$ (usually
$SU(N)$
or 
$SO(N)$);
$V_{U}$
commutes with
$U_{a,A}$ 
\item
$b_{J}, \quad J = 1,\dots,N_{B}$ (resp. $f_{A}, \quad A = 1,\dots,N_{F}$)
are the quantum free fields of integer (resp. half-integer) spin. We assume
that the fields are linearly independent up to equations of motion;
\item
The  equations of motion do not connect distinct fields. 
\end{itemize}

In this paper, we considers only particles of spin
$s \leq 1$.
All fields verify Klein-Gordon equation; if the Fermionic fields
are Majorana they verify Dirac equation.

Now we define the notion of extended supersymmetry invariance of the system of 
Bosonic and Fermionic fields considered above. Suppose that in the Hilbert 
space 
${\cal H}$ 
we also have the operators
$Q_{ja}, \quad j = 1,\dots,N, \quad a = 1,2$
such that:

(i) the following relations are verified:
\be
Q_{ja} \Omega = 0, \quad \bar{Q}_{j\bar{a}} \Omega = 0
\label{vac}
\ee
\bea
~[ Q_{ja}, P_{\mu} ] = 0, \quad
U_{A}^{-1} Q_{ja} U_{A} = {A_{a}}^{b} Q_{jb}, \quad \forall A \in SL(2,\C)
\nonumber \\
V_{U}^{-1} Q_{ja} V_{U} = \rho(U)_{jk} Q_{ka}, \quad \forall U \in G
\eea
and
\bea
\{ Q_{ja} , Q_{kb} \} = \epsilon_{ab} Z_{jk}, \quad
\{ Q_{ja} , \bar{Q}_{k\bar{b}} \} = 2 \delta_{jk}\sigma^{\mu}_{a\bar{b}} 
P_{\mu}, 
\label{SUSY}
\eea

Here
$P_{\mu}$
are the infinitesimal generators of the translation group given by
\be
[P_{\mu}, b ] = - i~\partial_{\mu} b, \quad
[P_{\mu}, f ] = - i~\partial_{\mu} f,
\label{P}
\ee
\be
\bar{Q}_{j\bar{b}} \equiv (Q_{jb})^{\dagger},
\ee
$Z_{jk}$
are the so-called {\it central charges} and, by definition, they commute with
all other SUSY generators and
$U \mapsto \rho(U)$ 
is $N$-dimensional a representation of the group $G$. We will consider only the
case
$\rho = Id$.    

(ii) The following commutation relations are true:
\bea
i [ Q_{ja}, b ] = p_{j}(\partial) f
\qquad
\{ Q_{ja}, f \} = q_{j}(\partial) b
\nonumber \\
i [ Z_{jk}, b ] = p_{jk}(\partial) b
\qquad
i [ Z_{jk}, f ] = q_{jk}(\partial) f.
\label{tensor}
\eea

Above $b$ (resp. $f$) is the collection of all integer (resp. half-integer) 
spin fields and $p, q$ are matrix-valued polynomials in the partial derivatives
$\partial_{\mu}$
(with constant coefficients). These relations express the tensor properties 
of the fields with respect to (infinitesimal) supersymmetry transformations. 

If this conditions are true we say that
$Q_{ja}$
are {\it super-charges} and
$b, f$
are forming a {\it supersymmetric multiplet}. The notion of 
{\it irreducibility} can be defined for any supersymmetric multiplet if
we consider the quantum fields as a modulus over the ring of partial
differential operators. As emphasized in \cite{GS1}, the matrix-valued 
operators
$p$ and $q$
are subject to some constraints which generalize the case
$N = 1$
(see \cite{GS1} and \cite{GS2}). 
\begin{itemize}
\item
From the compatibility of (\ref{tensor}) with Lorentz transformations it
follows that these polynomials are Lorentz covariant.
\item
The equation of motion are supersymmetric invariant, i.e. if we take the 
commutator of the supercharges
$Q_{ja}$
and
$\bar{Q}_{k\bar{a}}$
with the equations for the Bosonic fields we obtain zero modulo the equation
of motion for the Fermionic fields and vice-versa.
\item
To verify the validity of (\ref{SUSY}) it is necessary and sufficient to prove
they commute with all the fields 
$b$ and $f$
of the model: 
\bea
~[ \{ Q_{ja} , Q_{kb} \} - \epsilon_{ab} Z_{jk}, b ] = 0, \quad
[ \{ Q_{a} , \bar{Q}_{\bar{b}} \} - 2 \sigma^{\mu}_{a\bar{b}} P_{\mu} , b ] = 0
\nonumber \\
~[ \{ Q_{a} , Q_{b} \} - \epsilon_{ab} Z_{jk}, f ] = 0, \quad
[ \{ Q_{a} , \bar{Q}_{\bar{b}} \} - 2 \sigma^{\mu}_{a\bar{b}} P_{\mu} , f]  = 0;
\eea
this follows from (\ref{vac}) and the fact that the Hilbert space is generated 
by vectors of the type
\bea
\Psi = b_{J_{1}}(x_{1}) \cdots b_{J_{p}}(x_{p}) 
f_{A_{1}}(y_{1}) \cdots f_{A_{q}}(y_{q}) \Omega \in {\cal H}.
\label{hilbert}
\eea

Using the (graded) Jacobi identities it follows that we must check:
\bea  
\left\{ Q_{ja} , [ Q_{kb}, b ] \right\}  
- \left\{ Q_{kb} , [ b, Q_{ja} ] \right\} 
+ [ b , \epsilon_{ab} Z_{jk} ] = 0
\nonumber \\
\left[ Q_{ja} , \{ Q_{kb}, f \} \right] 
+ \left[ Q_{kb} , \{ f, Q_{ja} \} \right] 
+ [ f, \epsilon_{ab} Z_{jk} ] = 0
\nonumber \\
\left\{ Q_{ja} , [ \bar{Q}_{\bar{kb}}, b ] \right\} 
- \{ \bar{Q}_{\bar{kb}} , [ b, Q_{ja} ] \} 
+ [ b, 2 \delta{jk} \sigma^{\mu}_{a\bar{b}} P_{\mu} ] = 0
\nonumber \\
\left[ Q_{ja} , \{ \bar{Q}_{\bar{kb}}, f \} \right] 
+ [ \bar{Q}_{k\bar{b}} , \{ f, Q_{ja} \} ] 
+ [ f, 2 \delta{jk} \sigma^{\mu}_{a\bar{b}} P_{\mu} ] = 0.
\label{jacobi1}
\eea
\item
Causal (anti)commutation relations are verified by the free fields
$b$ and $f$:
\bea
~[ b(x), b(y) ] = - i p(\partial) D(x-y) \qquad
\{ f(x), f(y) \} = - i q(\partial) D(x-y)
\eea
where $D$ is Pauli-Jordan causal distribution. Reasoning as above it follows
that new consistency relations are valid following again from (graded) Jacobi 
identities; the non-trivial ones are:
\bea
[ b(x), \{ f(y), Q_{ja} \} ] + \{ f(y), [ Q_{ja}, b(x) ] \} = 0
\label{jacobi2}
\eea
\item
If one considers higher-spin fields (more precisely
$s \geq 1$),
is necessary to extend 
somewhat this framework: one considers in 
${\cal H}$
besides the usual positive definite scalar product a non-degenerate 
sesqui-linear form
$<\cdot, \cdot>$
which becomes positively defined when restricted to a factor Hilbert space
$Ker(Q)/Im(Q)$
where $Q$ is some {\it gauge charge}. We denote with
$A^{\dagger}$
the adjoint of the operator $A$ with respect to
$<\cdot, \cdot>$.

The gauge supercharge $Q$ it is usually determined by relations of the type 
(\ref{tensor}) involving ghost fields also, so it means that we must impose 
consistency relations of the same type as above. Moreover, it is desirable to 
have
\be
\{ Q, Q_{ja} \} = 0, \quad \{ Q, \bar{Q}_{j\bar{a}} \} = 0;
\label{susy-gauge}
\ee
this implies that the supersymmetric charges
$Q_{ja}$
and
$\bar{Q}_{j\bar{a}}$
factorizes to the physical Hilbert space
${\cal H}_{phys} = Ker(Q)/Im(Q)$.
This implies new consistency relations of the type (\ref{jacobi1})
with one of the supercharges replaced by the gauge charge:
\bea  
~\left\{ Q_{ja} , [ Q, b ] \right\} + \left\{ Q , [ Q_{ja}, b ] \right\} = 0
\qquad
~\left[ Q_{ja} , \{ Q, f \} \right] + \left[ Q, \{ Q_{ja}, f \} \right] = 0.
\label{jacobi3}
\eea
\item
A relation of the type (\ref{jacobi3}) must be also valid for the 
gauge charge:
\bea
[ b(x), \{ f(y), Q \} ] + \{ f(y), [ {Q}, b(x) ] \} = 0.
\label{jacobi4}
\eea
\item
To have
$Q^{2} = 0$
we must also impose
\be
~\{ Q, [Q , b ] \} = 0 \qquad [ Q, \{ Q, f \} ] = 0.
\label{jacobi5}
\ee
\end{itemize}

In the presence of a gauge charge one can relax (\ref{jacobi1}): Indeed one
can factorize the supercharges and the representation of the Poincar\'e group
to the physical Hilbert space
${\cal H}_{phys} = Ker(Q)/Im(Q)$
and require the (\ref{jacobi1}) are valid only for these factorized operators
\cite{GS2}.

\newpage

\section{$N = 2$ with the internal symmetry group $SU(2)$
and without central charges\label{n2}}

\begin{thm}
Let us consider the multiplet 
$(z, D, F_{ab}, C_{jk}, \lambda_{ja}, \chi_{ja}), \quad j = 1,2$
where:
\begin{itemize}
\item
$z, D$
are two complex Bosonic scalar field which are 
$SU(2)$
scalars; 
\item
$F_{ab}$
are complex Bosonic fields which are
$SU(2)$ 
scalars and such that
\be
F_{ab} = F_{ba};
\ee
\item
$C_{jk}$
are complex Bosonic fields which verify
\be
V_{U}^{-1} C_{jk} V_{U} = U_{jl} U_{km} C_{lm}, \quad \forall U \in SU(2)
\ee
and such that
\be
C_{jk} = C_{kj};
\ee
\item
$\lambda_{ja},\quad \chi_{ja}$
are Fermionic Dirac spinor fields and verifying
\be
V_{U}^{-1} \lambda_{ja} V_{U} = U_{jk} \lambda_{k}, \quad
V_{U}^{-1} \chi_{ja} V_{U} = U_{jk} \chi_{ka}, \quad \forall U \in SU(2).
\ee
\end{itemize}

The action of the supercharges on these fields is well defined through:
\be
i [ Q_{ja}, z ] = \chi_{ja} \qquad
[ \bar{Q}_{j\bar{a}}, z ] = 0.
\ee
\be
~[ Q_{ja} , F_{bc} ] = \epsilon_{ab} \lambda_{jc} + \epsilon_{ac} \lambda_{jb}
\ee
\be
~[ \bar{Q}_{j\bar{a}}, F_{bc} ] = - i \epsilon_{jk} 
(\sigma^{\mu}_{c\bar{a}} \partial_{\mu}\chi_{kb}
+ \sigma^{\mu}_{b\bar{a}} \partial_{\mu}\chi_{kc})
\ee
\be
~[ Q_{ja} , C_{kl} ] = 
- (\epsilon_{jk} \lambda_{la} + \epsilon_{jl} \lambda_{ka})
\ee
\be
~[ \bar{Q}_{j\bar{a}}, C_{kl} ] = i \sigma^{\mu}_{b\bar{a}}\partial_{\mu}
( \delta_{jk} \chi_{l}^{b} + \delta_{jl} \chi_{k}^{b})
\ee
\be
~[ Q_{ja} , D ] = 0
\ee
\be
[ \bar{Q}_{j\bar{a}}, D ] = 
2i \epsilon_{jk}\sigma^{\mu}_{b\bar{a}} \partial_{\mu}\lambda_{k}^{b}
\ee
\be
\{ Q_{ja}, \chi_{kb} \} = \epsilon_{jk} F_{ab} + \epsilon_{ab} C_{jk}
\ee
\be
\{ Q_{ja}, \bar{\chi}_{k\bar{b}} \} = 2 \delta_{jk} \sigma^{\mu}_{a\bar{b}}
\partial_{\mu}z^{*}
\ee
\be
\{ Q_{ja}, \lambda_{kb} \} = \epsilon_{jk} \epsilon_{ab} D
\ee
\be
\{ \bar{Q}_{j\bar{b}}, \lambda_{ka} \} = 
- i ( \delta_{jk} \sigma^{\mu}_{c\bar{b}} \epsilon^{cd}\partial_{\mu}F_{ad}
+ \epsilon_{jl} \sigma^{\mu}_{a\bar{b}} \partial_{\mu}C_{lk}).
\ee

If the field $z$ verifies the Klein-Gordon equation for mass $m$ then all
fields of the multiplet verify the same equation. 
\label{n=2}
\end{thm}
{\bf Proof:}
One can start from the existence of the fields $z$ and 
$\chi_{kb}$
derive the other fields of the multiplet from the Jacobi identities 
(\ref{jacobi1}). 
\begin{enumerate}
\item
From the relation involving;
$Q_{ja}, Q_{kb}$
and $z$ we get that the expression
$\{ Q_{ja}, \chi_{kb} \}$
is antisymmetric in the couples
$ja$ 
and 
$kb$ 
so if we define
\be
F_{ab} \equiv {1\over 2} \epsilon_{jk} \{ Q_{ja}, \chi_{kb} \}, \qquad
C_{jk} \equiv {1\over 2} \epsilon^{ba} \{ Q_{ja}, \chi_{kb} \}
\ee
we obtain the action of the supercharge
$Q_{ja}$
on
$\chi_{kb}$.
\item
The relation involving
$Q_{ja}, Q_{kb}$
and 
$z^{*}$
is an identity.
\item
The relation involving
$Q_{ja}, \bar{Q}_{k\bar{b}}$
and $z$
gives the action of the supercharge
$\bar{Q}_{k\bar{b}}$
on
$\chi_{ja}$.
\item
The relation involving
$Q_{ja}, \bar{Q}_{k\bar{b}}$
and 
$\chi_{lc}$
gives the action of the supercharge
$\bar{Q}_{k\bar{b}}$
on
$F_{ac}$
and
$C_{jk}$.
\item
The relation involving
$Q_{ja}, Q_{kb}$
and 
$\bar{\chi}_{l\bar{c}}$
is an identity.
\item
The relation involving
$Q_{ja}, Q_{kb}$
and 
$\chi_{lc}$
leads us to the introduction of a new field
\be
\lambda_{ja} \equiv - {1\over 3} \epsilon_{kl} [Q_{ka}, C_{jl} ]
\ee
and the action of the supercharge
$Q_{ka}$
on
$F_{bc}$
and
$C_{jl}$
follows.
\item
The relation involving
$Q_{ja}, \bar{Q}_{k\bar{b}}$
and 
$C_{lm}$
give the action of
$\bar{Q}_{k\bar{b}}$
on
$\lambda_{ma}$.
\item
The relation involving
$Q_{ja}, Q_{kb}$
and 
$C_{lm}$
leads us to the introduction of a new field
\be
d_{ab} \equiv \epsilon_{kl} \{ Q_{kb}, \lambda_{la} \};
\ee
one finds out the this expression must be antisymmetric in $a$ and $b$ so in
fact the new field is the (complex) scalar field $D$ such that:
\be
d_{ab} = {1\over 2} \epsilon_{ab} D;
\ee
we now have the action of
$Q_{ja}$
on
$\lambda_{kb}$.
\item
The relation involving
$\bar{Q}_{j\bar{a}}, \bar{Q}_{k\bar{b}}$
and 
$C_{lm}$
is an identity.
\item
The relation involving
$Q_{ja}$
(or
$\bar{Q}_{j\bar{a}}$),
$Q_{kb}$
(or
$\bar{Q}_{k\bar{b}}$)
and 
$F_{cd}$
are identities.
\item
The relation involving
$Q_{ja}, \bar{Q}_{k\bar{b}}$
and 
$\lambda_{lc}$
gives the action of
$\bar{Q}_{k\bar{b}}$
on $D$.
\item
The relation involving
$Q_{ja}, Q_{kb}$
and 
$\lambda_{lc}$
gives the action of
$Q_{kb}$
on $D$.
\item
The relation involving
$\bar{Q}_{j\bar{a}}, \bar{Q}_{k\bar{b}}$
and 
$\lambda_{lc}$
is an identity.
\item
The relation involving
$Q_{ja}$
(or
$\bar{Q}_{j\bar{a}}$),
$Q_{kb}$
(or
$\bar{Q}_{k\bar{b}}$) 
and 
$D$
are identities. 
\end{enumerate}
Finally one notices the compatibility with the $SU(2)$ transformations of
all the relations from the statement.
$\qed$

It is convenient to replace
$F_{ab}$
by new fields; we proceed in two stages with two elementary proposition.
\begin{prop}
We can write uniquely
\be
F_{ab} = \sigma^{\mu\nu}_{ab} {\cal F}_{\mu\nu}
\ee
if we impose antisymmetry and the duality condition
\be
{\cal F}_{\mu\nu} = {\cal F}_{\nu\mu}, \qquad
\epsilon^{\mu\nu\alpha\beta} {\cal F}_{\alpha\beta} = {\cal F}^{\nu\mu}. 
\ee

Then the action of the supercharges on the new fields is:
\be
~[ Q_{ja} , {\cal F}_{\mu\nu} ] = \sigma^{\mu\nu}_{ab} \lambda_{j}^{b}
\ee
\be
~[ \bar{Q}_{j\bar{a}}, {\cal F}_{\mu\nu} ] = {1\over 2} \epsilon_{jk} 
(\sigma^{\mu}_{b\bar{a}} \partial^{\nu}\chi_{k}^{b}
- \sigma^{\nu}_{b\bar{a}} \partial^{\mu}\chi_{k}^{b}
+ i \epsilon^{\mu\nu\rho\alpha} \sigma_{\alpha b\bar{a}} 
\partial_{\rho}\chi_{k}^{b}).
\ee
\end{prop}
Indeed, then we have:
\be
{\cal F}_{\mu\nu} = - {1\over 2} \sigma^{\mu\nu}_{ab} F^{ab}.
\ee

\begin{prop}
Let us define the real fields:
\be
F_{\mu\nu} \equiv {i \over 2} ( {\cal F}_{\mu\nu} - {\cal F}_{\mu\nu}^{*});
\ee
then the correspondence
$F_{ab} \leftrightarrow F_{\mu\nu}$
is one-one and the action of the supercharges on the new fields is:
\be
~[ Q_{ja}, F^{\mu\nu} ] = {i \over 2} \sigma^{\mu\nu}_{ab} \lambda_{j}^{b}
+ {i\over 4} \epsilon_{jk} 
(\sigma^{\mu}_{a\bar{b}} \partial^{\nu}\bar{\chi}_{k}^{\bar{b}}
- \sigma^{\nu}_{a\bar{b}} \partial^{\mu}\bar{\chi}_{k}^{\bar{b}}
- i \epsilon^{\mu\nu\rho\alpha} \sigma_{\alpha a\bar{b}} 
\partial_{\rho}\bar{\chi}_{k}^{\bar{b}}).
\ee
\end{prop}

Indeed the correspondence is one-one because we have
\be
{\cal F}_{\mu\nu} = F_{\mu\nu} - {i \over 2} \epsilon_{\mu\nu\alpha\beta}
F^{\alpha\beta}.
\ee

The preceding multiplet is reducible. 
\begin{thm}
The condition
\be
C_{jk}^{*} = \epsilon_{jl} \epsilon_{km} C_{lm}
\label{constr1}
\ee
is compatible with the $SU(2)$ transformations. It is consistent with the
action of the supercharges from the preceding theorem {\it iff} we also have:
\be
\lambda_{ja} = i \epsilon_{jk} \sigma^{\mu}_{a\bar{b}} 
\partial_{\mu}\bar{\chi}_{k}^{\bar{b}}
\label{constr2}
\ee
\be
D = - 2 i \partial^{2}z^{*}
\label{constr3}
\ee
\be
\epsilon^{\mu\nu\alpha\beta}\partial_{\nu}F_{\alpha\beta} = 0.
\label{constr4}
\ee
\end{thm}
{\bf Proof:}
One starts from the first constrain and by commuting with the supercharges
gets new constraints. Then we one iterates the procedure and, hopefully, it
will close after a finite number of steps. 
\begin{enumerate}
\item
We commute the constrain (\ref{constr1}) with the supercharge
$Q_{ja}$
and obtain (\ref{constr2}).
\item
The commutator of (\ref{constr1}) with the supercharge
$\bar{Q}_{k\bar{b}}$
is an identity.
\item
We anticommute the constraint (\ref{constr2}) with the supercharge
$Q_{ja}$
and obtain (\ref{constr3}).
\item
We anticommutator the constraint (\ref{constr2}) with the supercharge
$\bar{Q}_{k\bar{b}}$
and we obtain (\ref{constr4}).
\item
The commutator of (\ref{constr3}) and (\ref{constr4}) with the supercharges
$Q_{ja}, \bar{Q}_{k\bar{b}}$
are identities if we use the preceding constraints.
\end{enumerate}
$\qed$

It follows that the reduced multiplet can be described as follows:
\begin{thm}
The reduced multiplet is composed of the fields
$(z,F_{\mu\nu},C_{jk},\chi_{ja})$
verifying the restrictions
\be
C_{jk} = C_{kj}, \qquad C_{jk}^{*} = \epsilon_{jl} \epsilon_{km} C_{lm}
\ee
\be
F_{\mu\nu} = - F_{\nu\mu}, \qquad
\epsilon^{\mu\nu\alpha\beta}\partial_{\nu}F_{\alpha\beta} = 0.
\ee

The action of the supercharges on these fields is well defined by:
\be
i [ Q_{ja}, z ] = \chi_{ja} \qquad
[ \bar{Q}_{j\bar{a}}, z ] = 0.
\ee
\be
~[ Q_{ja}, F^{\mu\nu} ] = {i\over 2} \epsilon_{jk} 
(\sigma^{\mu}_{a\bar{b}} \partial^{\nu}\bar{\chi}_{k}^{\bar{b}}
- \sigma^{\nu}_{a\bar{b}} \partial^{\mu}\bar{\chi}_{k}^{\bar{b}})
\ee
\be
~[ Q_{ja} , C_{kl} ] = - i \sigma^{\mu}_{a\bar{b}}\partial_{\mu}
(\epsilon_{jk} \epsilon_{lm} \bar{\chi}_{m}^{\bar{b}} 
+ \epsilon_{jl} \epsilon_{km}\bar{\chi}_{m}^{\bar{b}})
\ee
\be
~[ \bar{Q}_{j\bar{a}}, C_{kl} ] = i \sigma^{\mu}_{b\bar{a}}\partial_{\mu}
( \delta_{jk} \chi_{l}^{b} + \delta_{jl} \chi_{k}^{b})
\ee
\be
\{ Q_{ja}, \chi_{kb} \} = - 2 i \epsilon_{jk} \sigma^{\mu\nu}_{ab} F_{\mu\nu} 
+ \epsilon_{ab} C_{jk}
\ee
\be
\{ Q_{ja}, \bar{\chi}_{k\bar{b}} \} = 2 \delta_{jk} \sigma^{\mu}_{a\bar{b}}
\partial_{\mu}z^{*}.
\ee
\label{reduced}
\end{thm}

The preceding multiplet can be reduced even further. We have:
\begin{thm}
In the preceding conditions the relations
\be
C_{jk} = 0
\ee
\be
\sigma^{\mu}_{a\bar{b}}\partial_{\mu}\bar{\chi}_{j}^{\bar{b}} = 0
\ee
\be
\partial^{2}z = 0
\ee
\be
\partial^{\mu}F_{\mu\nu} = 0
\ee
are supersymmetric invariant. In particular we can consider the multiplet
$(z,\chi_{ja},F_{\mu\nu})$
of zero mass fields such that
$\chi_{ja}$
verifies Dirac equation of zero mass and
$F_{\mu\nu}$
verifies the restrictions
\be
F_{\mu\nu} = - F_{\nu\mu}, \qquad
\epsilon^{\mu\nu\alpha\beta}\partial_{\nu}F_{\alpha\beta} = 0 \qquad
\partial^{\mu}F_{\mu\nu} = 0.
\ee

The action of the supercharges is well defined by:
\be
i [ Q_{ja}, z ] = \chi_{ja} \qquad
[ \bar{Q}_{j\bar{a}}, z ] = 0.
\ee
\be
~[ Q_{ja}, F^{\mu\nu} ] = {i\over 2} \epsilon_{jk} 
(\sigma^{\mu}_{a\bar{b}} \partial^{\nu}\bar{\chi}_{k}^{\bar{b}}
- \sigma^{\nu}_{a\bar{b}} \partial^{\mu}\bar{\chi}_{k}^{\bar{b}})
\ee
\be
\{ Q_{ja}, \chi_{kb} \} = - 2 i \epsilon_{jk} \sigma^{\mu\nu}_{ab} F_{\mu\nu} 
\ee
\be
\{ Q_{ja}, \bar{\chi}_{k\bar{b}} \} = 2 \delta_{jk} \sigma^{\mu}_{a\bar{b}}
\partial_{\mu}z^{*}.
\ee
\label{reduced1}
\end{thm}

We call this multiplet the {\it new reduced multiplet}.

We can make now the connection with the notations of \cite{Fa}. We have to
show that the fields
$C_{jk}$
are transforming according to the vector representation of
$SU(2)$.
Indeed we have
\begin{lemma}
(i) In the preceding conditions, let us define the operator-valued 
$2 \times 2$
matrix $H$ with elements
\be
H_{jk} \equiv C_{jl}~\epsilon_{lk}.
\ee

Then we have
\be
H^{T} = \epsilon~H~\epsilon
\label{anti}
\ee
\be
V^{-1}_{U}~H~V_{U} = U~H~U^{*}, \quad \forall U \in SU(2).
\label{h}
\ee

(ii) Any 
$2 \times 2$
matrix $H$ verifying the relation (\ref{anti}) can be uniquely written in the 
form
\be
H = \sum_{j=1}^{3} C_{j} \sigma_{j}
\label{vec-h}
\ee
(where
$\sigma_{j}$
are the Pauli matrices) so (\ref{h}) is equivalent to
\be
V^{-1}_{U}~C_{j}~V_{U} = \delta(U)_{jk}~C_{k}.
\label{vec-c}
\ee

Here 
$\delta:SU(2) \rightarrow SO(3)$
is the covering map.
\end{lemma}
{\bf Proof:}
The relations (i) are elementary and the writing (\ref{vec-h}) follows from
$$
\sigma_{j}^{T} = \epsilon~\sigma_{j}~\epsilon, \quad \forall j = 1,2,3.
$$
The relation (\ref{vec-c}) tells us that the multiplet of fields
$C_{j}$
transforms according to the vector representation of
$SU(2)$.
$\qed$

Now the multiplet described in \cite{Fa} by the rules (12)+(16) coincides with
the multiplet from our theorem \ref{reduced}: indeed, up to some factors
we have the correspondence:
\bea
z = A + i B, \quad \chi_{1a} = p_{a}, \quad \chi_{2a} = p_{a}, 
\quad F^{\mu\nu} = V^{\mu\nu} \quad
Q_{1a} = Q_{\uparrow a}, \quad Q_{2a} = Q_{\downarrow a}.
\nonumber
\eea  

The new reduced multiplet corresponds to (18) from \cite{Fa}. Let us
remember that the action of the supercharges on the fields described in
theorem \ref{n=2} implies that we do have (\ref{SUSY}) without central
charges;  in fact, we have derived the action of the supercharges
imposing this conditions, the relations (\ref{SUSY}) being generic for a
supersymmetry algebra. A posteriori we have (\ref{SUSY}) for
theorems \ref{reduced} and \ref{reduced1}. In \cite{Fa} it is asserted that
the anticommutation relations between supercharges corresponding to 
$j = 1$ 
and  
$j = 2$  
are unconstrained; according to our analysis this is not true.

Now we determine the possible form of the causal (anti)commutator relations.
We have the following result:
\begin{thm}
Let us suppose that for the multiplet of Theorem \ref{n=2} the field $z$ is a 
complex scalar field of mass 
$m \geq 0$;
in particular we have
\be
~[ z(x), z(y) ] = 0
\ee
\be
~[ z(x), z^{*}(y) ] = - i~D_{m}(x-y).
\ee

Then we also have:
\be
~[ F_{ab}(x), F_{cd}(y) ] = i~\alpha~
(\epsilon_{ac} \epsilon_{bd} + \epsilon_{ad} \epsilon_{bc}) D_{m}(x-y)
\ee
\be
~[ F_{ab}(x), F_{\bar{c}\bar{d}}^{*}(y) ] = 
2 i~\left( \sigma^{\mu}_{a\bar{c}} \sigma^{\nu}_{b\bar{d}}
+ \sigma^{\mu}_{a\bar{d}} \sigma^{\nu}_{b\bar{c}}\right)~
\partial_{\mu}\partial_{\nu}D_{m}(x-y)
\ee
\be
~[ C_{jk}(x), C_{lm}(y) ] = - i~\alpha 
(\epsilon_{jl} \epsilon_{km} + \epsilon_{jm} \epsilon_{kl}) D_{m}(x-y)
\ee
\be
~[ C_{jk}(x), C_{lm}^{*}(y) ] = - 2~i~m^{2}
(\delta_{jl} \delta_{km} + \delta_{jm} \delta_{kl}) D_{m}(x-y)
\ee
\be
~[ D(x), D^{*}(y) ] = - 4~im^{2}~D_{m}(x-y)
\ee
\be
~[ z(x), D(y) ] = \alpha~D_{m}(x-y)
\ee
\be
\left\{ \chi_{ja}(x), \bar{\chi}_{k\bar{b}}(y) \right\} =
2~\delta_{jk} \sigma^{\mu}_{a\bar{b}}~\partial_{\mu}D_{m}(x-y)
\ee
\be
\left\{ \lambda_{ja}(x), \bar{\lambda}_{k\bar{b}}(y) \right\} =
2~\delta_{jk} m^{2}~\sigma^{\mu}_{a\bar{b}}~\partial_{\mu}D_{m}(x-y)
\ee
\be
\left\{ \chi_{ja}(x), \lambda_{kb}(y) \right\} 
= - i \alpha~\epsilon_{ab}~\epsilon_{jk}~D_{m}(x-y),
\ee
and all other (anti)commutators are zero. Here
$\alpha \in \C$
is a free parameter. 
\end{thm}
{\bf Proof:}
From Lorentz covariance considerations we must have:
\be
~[ z(x), F_{ab}(y) ] = 0, \qquad
~[ z^{*}(x), F_{ab}(y) ] = 0
\ee
because there is no partial differential operator
$P_{ab}(\partial)$
symmetric in $a$ and $b$ and Lorentz covariant. We also must have
\be
~[ z(x), D(y) ] = \alpha~D_{m}(x-y), \qquad
~[ z(x), D^{*}(y) ] = \alpha^{\prime}~D_{m}(x-y)
\ee
with 
$\alpha, \alpha^{\prime} \in \C$.

From consideration of Lorentz and $SU(2)$ covariance we must also have:
\be
~[ z(x), C_{jk}(y) ] = 0
\ee
because the only available combination
$
\epsilon_{jk} D_{m}(x-y)
$
is in conflict with the symmetry property.

Starting from the hypothesis and the preceding relations one determines by some 
long but straightforward computation all the causal (anti)commutators from the
Jacobi identity (\ref{jacobi2}). In particular we get
$\alpha^{\prime} = 0$.
$\qed$

Now we determine if the multiplets considered above do admit a representation
in a Fock space. We proceed in the spirit of the reconstruction theorem from
the axiomatic field theory. The key property is positivity. We suppose that
all the fields are free of mass $m$. In this case it is known that it is
sufficient to verify the positivity of the $2$-point function. We have
immediately
\begin{cor}
The field $D$ has a representation in a Fock space iff 
$m > 0$
\end{cor}
{\bf Proof:}
Let us consider the real scalar fields
$D_{r}, r = 1,2$
given by 
$D = D_{1} + i~D_{2}$.
From the commutation relation for $D$ we get:
\be
[ D_{r}(x), D_{s}(y) ] = - 2i~m^{4}~\delta_{rs}~D_{m}(x-y) 
\ee
so the $2$-points functions is obtained from the causal commutator with
the substitution
\be
D_{m} \rightarrow D_{m}^{(+)}
\label{subst}
\ee
i.e. we keep only the positive frequency part of the Pauli-Villars commutator:
\be
<\Omega, D_{r}(x) D_{s}(y) \Omega> = - 2~im^{2}~\delta_{rs}~D^{(+)}_{m}(x-y);
\ee
this follows from K\"allan-Lehman representation for the 2-point distribution
(see for instance \cite{IAS}, {\bf Introduction to QFT}, Section 1.5).
If we consider the norm of an arbitrary one-particle state we get something
proportional to 
$m^{4}$
so the positivity condition follows.
$\qed$

To analyse the possibility of Hilbert space representations for the other
field more easily we rewrite some of the commutation  relations given
above using new fields. The first rewriting is elementary in  terms of
the fields 
$F_{\mu\nu}$.
\begin{cor}
The causal commutation relations for the field
$F_{\mu\nu}$
are
\bea
[ F_{\mu\nu}(x), F_{\rho\sigma}(y) ] = - {i\over 2}
( g_{\mu\rho} \partial_{\nu}\partial_{\sigma}
- g_{\nu\rho} \partial_{\mu}\partial_{\sigma}
+ g_{\nu\sigma} \partial_{\mu}\partial_{\rho}
- g_{\nu\rho} \partial_{\mu}\partial_{\sigma})~D_{m}(x-y)
\nonumber \\
+ {i\over 8} [ Re(\alpha) - 2 m^{2} ] 
( g_{\mu\rho} g_{\nu\sigma} - g_{\mu\sigma} g_{\nu\rho} )~D_{m}(x-y)
+ {i\over 8} Im(\alpha) \epsilon_{\mu\nu\rho\sigma}~D_{m}(x-y). 
\eea

In particular the field has a representation in a Fock space if and only if
$\alpha = 2 m^{2}$. 
\label{pos1}
\end{cor}
{\bf Proof:}
The commutation relation is obtained by a straightforward computation. The 
$2$-point function is again obtained from the causal commutator with the
substitution (\ref{subst}). The computation of the norm of an arbitrary
one-particle state 
\be
\Psi \equiv \int dx~f^{\mu\nu}(x) F_{\mu\nu}(x) \Omega
\ee
where
$f^{\mu\nu}$
are test function (antisymmetric in $\mu$ and $\nu$) gives
\bea
|\Psi|^{2} = - 2\pi \int d\alpha^{+}_{m}(p) 
\overline{\tilde{f}}^{\mu}(p) \tilde{f}_{\mu}(p)
\nonumber \\
+ {\pi\over 2} [ Re(\alpha) - 2 m^{2} ] \int d\alpha^{+}_{m}(p) 
\overline{\tilde{f}}^{\mu\nu}(p) \tilde{f}_{\mu\nu}(p)
- {\pi\over 4} Im(\alpha)  \int d\alpha^{+}_{m}(p) \epsilon^{\mu\nu\rho\sigma}
\overline{\tilde{f}}_{\mu\nu}(p) \tilde{f}_{\rho\sigma}(p).
\eea

Here
$\alpha^{+}_{m}$
is the Lorentz-invariant measure on the upper hyperboloid of mass $m$,
$f^{\mu} \equiv \partial_{\nu}f^{\mu\nu}$
and by 
$\tilde{f}$
we denote the Fourier transform of $f$. Now it is clear that the last two
contributions can have arbitrary signs so they must be identically zero.
On the contrary, because of the transversality condition
$
\partial_{\mu}f^{\mu} = 0 \quad \Leftrightarrow p_{\mu} \tilde{f}(p) = 0
$
the first contribution is positive.
$\qed$

To analyze the positivity condition for the fields
$C_{jk}$
we proceed in the same spirit as for the complex field $D$: we make a 
decomposition into real and imaginary parts but in a $SU(2)$ covariant way.
\begin{cor}
Let us define the new fields
\be
{\cal C}_{jk} \equiv 
{1\over 2} (C_{jk} + \epsilon_{jl} \epsilon_{km} C_{km}^{*}), \qquad 
{\cal D}_{jk} \equiv 
{i\over 2} (C_{jk} - \epsilon_{jl} \epsilon_{km} C_{km}^{*}).
\ee

The correspondence
$C_{jk} \leftrightarrow ({\cal C}_{jk},{\cal D}_{jk})$
is one-one. These new fields verify the reality conditions
\be
{\cal C}^{*}_{jk} = \epsilon_{jl} \epsilon_{km} {\cal C}_{lm}, \qquad 
{\cal D}^{*}_{jk} = \epsilon_{jl} \epsilon_{km} {\cal D}_{lm} 
\ee
and the causal commutators are:
\be
[ {\cal C}_{jk}(x), {\cal C}^{*}_{lm}(y) ] = - {i \over 2} (2 m^{2} + \alpha)
(\delta_{jl} \delta_{km} + \delta_{jm} \delta_{kl}) D_{m}(x-y)
\ee
\be
[ {\cal D}_{jk}(x), {\cal D}^{*}_{lm}(y) ] = - {i \over 2} (2 m^{2} - \alpha )
(\delta_{jl} \delta_{km} + \delta_{jm} \delta_{kl}) D_{m}(x-y)
\ee
\be
[ {\cal C}_{jk}(x), {\cal D}_{lm}(y) ] = 0.
\ee

In particular the positivity condition is fulfilled iff
$- 2 m^{2} < \alpha < 2 m^{2}$.
\end{cor}
{\bf Proof:}
The commutators are obtained by elementary computation. If we consider now
the one-particle states
\be
\Psi_{1} \equiv \int dx~f_{jk}(x) {\cal C}^{*}_{jk}(x) \Omega, \qquad
\Psi_{2} \equiv \int dx~g_{jk}(x) {\cal D}^{*}_{jk}(x) \Omega
\ee
where
$f_{jk}, g_{jk}$
are test functions (antisymmetric in $j$ and $k$) then the norms are:
\bea
|\Psi_{1}|^{2} = 2\pi (2 m^{2} + \alpha) \int d\alpha^{+}_{m}(p) 
|{\tilde{f}}_{jk}(p)|^{2} 
\nonumber \\
|\Psi_{2}|^{2} = 2\pi (2 m^{2} - \alpha) \int d\alpha^{+}_{m}(p) 
|{\tilde{f}}_{jk}(p)|^{2} 
\eea
and we get the inequalities from the statement.
$\qed$

Comparing the last two corollaries we arrive at the conclusion that the
multiplet described in Theorem \ref{n=2} does {\bf not} have a representation
in a Hilbert space (of Fock type). This conclusion makes the reduced multiplet
more interesting because in this case we have positivity.
\begin{thm}
The causal (anti)commutators for the reduced multiplet of Theorem \ref{reduced}
are
\be
~[ z(x), z^{*}(y) ] = - i~D_{m}(x-y).
\ee
\be
[ F_{\mu\nu}(x), F_{\rho\sigma}(y) ] = - {i\over 2}
( g_{\mu\rho} \partial_{\nu}\partial_{\sigma}
- g_{\nu\rho} \partial_{\mu}\partial_{\sigma}
+ g_{\nu\sigma} \partial_{\mu}\partial_{\rho}
- g_{\nu\rho} \partial_{\mu}\partial_{\sigma})~D_{m}(x-y)
\ee
\be
~[ C_{jk}(x), C_{lm}(y) ] = - 2 i m^{2} 
(\epsilon_{jl} \epsilon_{km} + \epsilon_{jm} \epsilon_{kl}) D_{m}(x-y)
\ee
\be
\left\{ \chi_{ja}(x), \bar{\chi}_{k\bar{b}}(y) \right\} =
2~\delta_{jk} \sigma^{\mu}_{a\bar{b}}~\partial_{\mu}D_{m}(x-y).
\ee

This corresponds to
$\alpha = 2 m^{2}$;
in particular we have
${\cal D}_{jk} = 0$
so
${\cal C}_{jk} = C_{jk}.$
The positivity condition is verified in this case for
$m > 0$.
\end{thm}
{\bf Proof:}
The computations are elementary: from the constraints we have
$D = 2 i m^{2} z^{*}$;
if we substitute this into the commutation relation of $z$ with $D$
we get the value of $\alpha$. This in turn gives
${\cal D}_{jk} = 0$ 
and the commutators from the statement follows. If
$m = 0$
we would get that the fields
$C_{jk}$
are commuting among themselves so they cannot be represented as operators
in a Hilbert space. (They cannot be c-number fields because this would
contradict the relations from theorem \ref{n=2}).
$\qed$

Let us remark that in \cite{Fa} the reduced multiplet is constructed for
$m = 0$.
Apparently one can make 
$m = 0$
in theorem \ref{reduced} at the purely algebraic level and still obtain a good
multiplet. However, according to our analysis the multiplet obtained after 
this limiting procedure will not have a representation in a Hilbert space.

For the new reduced multiplet we have similarly:
\begin{thm}
The causal (anti)commutators for the reduced multiplet of Theorem \ref{reduced1}
are
\be
~[ z(x), z^{*}(y) ] = - i~D_{m}(x-y).
\ee
\be
[ F_{\mu\nu}(x), F_{\rho\sigma}(y) ] = - {i\over 4}
( g_{\mu\rho} \partial_{\nu}\partial_{\sigma}
- g_{\nu\rho} \partial_{\mu}\partial_{\sigma}
+ g_{\nu\sigma} \partial_{\mu}\partial_{\rho}
- g_{\nu\rho} \partial_{\mu}\partial_{\sigma})~D_{m}(x-y)
\ee
\be
\left\{ \chi_{ja}(x), \bar{\chi}_{k\bar{b}}(y) \right\} =
2~\delta_{jk} \sigma^{\mu}_{a\bar{b}}~\partial_{\mu}D_{m}(x-y).
\ee

The positivity condition is verified in this case.
\end{thm}

The reduced (and the new reduced) multiplet describe particles of spin $1$ 
through the fields
$F_{\mu\nu}$. 
Indeed it is easy to see that the one-particle Fock subspace
${\cal H}^{(1)}$
generated from the vacuum by
$F_{\mu\nu}$
describes the irreducible representation
$[m,1]$
of the Poincar\'e group. For the reduced multiplet we must have
$m > 0$
and for the new reduced multiplet we have
$m = 0$.

One can associate with every field of the multiplet from Theorem \ref{n=2}
a superfield using a sandwich formula as in \cite{GS1} and \cite{GS2}: if
$f$ is any field of the multiplet we define:
\be
s(f) \equiv e^{iS} f e^{-iS}
\ee 
where
\be
S \equiv \epsilon_{jk}(\theta^{a}_{j} Q_{ka} 
- \bar{\theta}^{\bar{a}}_{j} \bar{Q}_{k\bar{a}});
\ee
here
$\theta_{ja}$
are some arbitrary Grassmann parameters. The presence of the tensor
$\epsilon_{jk}$
makes the construction $SU(2)$-covariant. For instance, the superfield
$Z \equiv s(z)$
verifies the following covariance property 
$\forall U \in SU(2)$:
\be
V_{U}^{-1} Z(x,\theta) V_{U} = Z(x, U\cdot\theta) 
\ee
where we have defined the following action of $SU(2)$ on the Grassmann 
variables:
\be
(U\cdot\theta)_{ja} \equiv U_{jk} \theta_{ka}.
\ee

The superfield $Z$ is chiral i.e. we have
\be
\bar{\cal D}_{j\bar{a}} Z = 0
\ee
where the covariant derivatives are defined as usual \cite{GS1} for both 
values of the index $j$.

We also mention that the constraints (\ref{constr1})-(\ref{constr4}) can be
express compactly using the superfield $Z$
\be
\bar{\cal D}_{j\bar{a}} \bar{\cal D}_{k}^{\bar{a}} Z 
+ \epsilon_{jl} \epsilon_{km} {\cal D}_{l}^{a} {\cal D}_{ma} Z = 0.
\ee

The explicit expression for the superfield $Z$ is rather complicated and so is
the corresponding causal commutator.

Finally we investigate if it is possible to introduce in the game the
the electromagnetic potential
$A_{\mu}$.

Apparently this is possible for the reduced multiplet because the constraint
(\ref{constr4}) is the homogeneous Maxwell equation which tells us that
there exists
$A_{\mu}$
such that
\be
F_{\mu\nu} \equiv \partial_{\mu}A_{\nu} - \partial_{\nu}A_{\mu} 
\label{FA}
\ee

One needs the action of the supercharges on
$A_{\mu}$.
There are two distinct possibilities. The simplest one is to observe that the 
``lift" the action of the supercharges on given by
\be
~[ Q_{ja}, A^{\mu} ] = {i\over 2} \epsilon_{jk} 
\sigma^{\mu}_{a\bar{b}} \bar{\chi}_{k}^{\bar{b}}
\label{Q-A}
\ee
is compatible with (\ref{FA}) and the action on the supercharges from the 
Theorem \ref{reduced}. The closeness of the supersymmetric algebra is more 
subtle and can be understood as in \cite{GS2}, Sect. 8. However, if one 
computes the causal commutators one finds out that the fields
$C_{jk}$
causally commutes with every other fields (including themselves) so we do
not have a Hilbert space representation. The situation is better for the 
new reduced multiplet from Theorem \ref{reduced1} for which the non-trivial
causal (anti)commutators are:
\be
~[ z(x), z^{*}(y) ] = - i~D_{m}(x-y).
\ee
\be
[ F_{\mu\nu}(x), F_{\rho\sigma}(y) ] = - {i\over 4}
( g_{\mu\rho} \partial_{\nu}\partial_{\sigma}
- g_{\nu\rho} \partial_{\mu}\partial_{\sigma}
+ g_{\nu\sigma} \partial_{\mu}\partial_{\rho}
- g_{\nu\rho} \partial_{\mu}\partial_{\sigma})~D_{m}(x-y)
\ee
\be
\left\{ \chi_{ja}(x), \bar{\chi}_{k\bar{b}}(y) \right\} =
2~\delta_{jk} \sigma^{\mu}_{a\bar{b}}~\partial_{\mu}D_{m}(x-y)
\ee
so we do have a Hilbert space representations.

The second possibility of introducing the field
$A_{\mu}$
is more logical. First one observes that for 
$m > 0$
one can always transform an index $a$ into an index $\bar{a}$ and vice-versa
using the Dirac operator. In particular, if we start from the basic field
$F_{ab}$
of the multiplet appearing in Theorem \ref{n=2} we can define
\be
F_{a\bar{b}} \equiv \sigma^{\mu}_{c\bar{b}} \epsilon^{cd} 
\partial_{\mu}F_{ad}.
\ee 

One can check two facts: (a) The association
$F_{ab} \rightarrow F_{a\bar{b}}$
one-one. Indeed, one easily get from the preceding definition
\be
F_{ab} = {1\over m^{2}} \sigma^{\mu}_{b\bar{c}} \epsilon^{\bar{c}\bar{d}} 
\partial_{\mu}F_{a\bar{d}}
\ee 
which is the inverse of the map
$F_{ab} \rightarrow F_{a\bar{b}}$.
(b) The new field
$F_{a\bar{b}}$
verifies the property
\be
\sigma^{\mu}_{b\bar{c}} \epsilon^{\bar{c}\bar{d}} \partial_{\mu}F_{a\bar{d}}
= a \leftrightarrow b.
\label{sym1}
\ee 

So the first step is to replace in the multiplet
$F_{ab}$
by
$F_{a\bar{b}}$
One can give the action of the supercharges for the transformed multiplet:
\be
~[ Q_{ja} , F_{b\bar{c}} ] = \epsilon_{ab} \sigma^{\mu}_{d\bar{c}}
\partial_{\mu}\lambda_{j}^{d} - \sigma^{\mu}_{a\bar{c}} 
\partial_{\mu}\lambda_{jb}
\ee
\be
~[ \bar{Q}_{j\bar{a}}, F_{b\bar{c}} ] = - i \epsilon_{jk} 
(\sigma^{\mu}_{d\bar{c}} \sigma^{\rho}_{b\bar{a}}
\partial_{\mu}\partial_{\rho}\chi_{kb}
- m^{2}\epsilon_{\bar{a}\bar{c}} \chi_{kb}).
\ee

We easily prove that for the reduced multiplet from Theorem \ref{reduced} one
can impose the reality condition:
\be
(F_{a\bar{b}})^{*} =  F_{b\bar{a}}.
\ee

Next, one defines
\be
A_{\mu} \equiv {1\over 4 m^{2}} \sigma^{\mu}_{a\bar{b}} 
\epsilon^{ac} \epsilon^{\bar{b}\bar{d}} F_{b\bar{d}}. 
\ee

One can easily prove the following facts: (a) The association 
$F_{a\bar{b}} \rightarrow A_{\mu} $
is one-one. (b) The field
$A_{\mu}$
verifies the transversality condition
\be
\partial^{\mu}A_{\mu} = 0;
\ee
(this follows from (\ref{sym1})). (c) If the field
$F_{a\bar{b}}$
is real then the field
$A_{\mu}$
is also real and vice-versa.

This means that we can replace the (real) field
$F_{a\bar{b}}$
by the (real) field
$A_{\mu}$.
The action of the supercharges in this new representation is:
\be
~i~[ Q_{ja} , A_{\mu} ] = \epsilon_{ab} \sigma^{\rho}_{a\bar{b}}
\left( g_{\mu\rho} + {1\over m^{2}} \partial_{\mu}\partial_{\rho}\right)
\bar{\chi}_{k}^{\bar{b}}.
\ee

One can also compute the causal commutation relations for the new field
$A_{\mu}$
using the two successive transformations; one gets:
\be
[ A_{\mu}(x), A_{\rho}(y) ] = {i\over 8}~
\left( g_{\mu\rho} + {1\over m^{2}} \partial_{\mu}\partial_{\rho}\right)~
D_{m}(x-y).
\ee

The preceding two relations are compatible with the transversality property.
However one knows that in the usual formulation of the standard model one 
does not impose this transversality property \cite{Sc} so this multiplet
cannot be used for a supersymmetric extension of the standard model.

\newpage
\section{The $N = 2$ Hyper-multiplet\label{hypermultiplet}}

We consider now a multiplet with $SU(2)$ invariance and a non-trivial central
charge. It can be argued easily that we must have in (\ref{SUSY})
\be
Z_{jk} = \epsilon_{jk} Z
\ee
where $Z$ is a central charge and is a $SU(2)$ scalar. Moreover we can suppose
that it is self-adjoint
\be
Z^{*} = Z.
\ee

We have now
\begin{thm}
Let us consider the multiplet 
$(\phi_{j}, f_{j}, \psi_{a}, \chi_{a}), \quad j = 1,2$
where:
\begin{itemize}
\item
$\phi_{j}, f_{j}$
are complex Bosonic scalar field verify
\be
V_{U}^{-1} \phi_{j} V_{U} = U_{jk} \phi_{k}, \quad
V_{U}^{-1} f_{j} V_{U} = U_{jk} f_{k}, \quad
\quad \forall U \in SU(2).
\ee
\item
$\psi_{a},\quad \chi_{a}$
are Fermionic Dirac spinor fields which are $SU(2)$ scalars.
\end{itemize}

The action of the supercharges and of the central charge on these fields is 
well defined through:
\be
i [ Q_{ja}, \phi_{k} ] = \epsilon_{jk} \psi_{a}
\ee
\be
i [ \bar{Q}_{j\bar{a}}, \phi_{k} ] = \delta_{jk} \bar{\chi}_{\bar{a}}
\ee
\be
~[ Q_{ja} , f_{k} ] = i\epsilon_{jk} \sigma^{\mu}_{a\bar{b}}
\partial_{\mu}\bar{\chi}^{\bar{b}}
\ee
\be
~[ \bar{Q}_{j\bar{a}}, f_{k} ] = - i \delta_{jk} 
\sigma^{\mu}_{b\bar{a}} \partial_{\mu}\psi^{b}
\ee
\be
\{ Q_{ja}, \psi_{b} \} = - 2 \epsilon_{ab} f_{j}
\ee
\be
\{ Q_{ja}, \bar{\psi}_{\bar{b}} \} = 2 \epsilon_{jk} \sigma^{\mu}_{a\bar{b}}
\partial_{\mu}\phi_{k}^{*}
\ee
\be
\{ Q_{ja}, \chi_{b} \} = 2 \epsilon_{jk} \epsilon_{ab} f^{*}_{k}
\ee
\be
\{ Q_{ja}, \bar{\chi}_{\bar{b}} \} = 2 \sigma^{\mu}_{a\bar{b}}
\partial_{\mu}\phi_{j}
\ee
\be
i [ Z , \phi_{j} ] = f_{j}
\ee
\be
i [ Z, f_{j} ] = i \partial^{2} \phi_{j} 
\ee
\be
~[ Z , \psi_{a} ] = i~\sigma^{\mu}_{b\bar{a}}\partial_{\mu}\bar{\chi}^{\bar{b}}
\ee
\be
~[ Z, \chi_{a} ] = - i~\sigma^{\mu}_{a\bar{b}}\partial_{\mu}\bar{\psi}^{\bar{b}}
\ee

If the fields 
$\phi_{j}$ 
verifies the Klein-Gordon equation for mass $m$ then all fields of the 
multiplet verify the same equation. 
\label{hyper}
\end{thm}
{\bf Proof:}
One assume that the action of the supercharges and of the central charge on 
$\phi_{j}$ 
is given by the formul\ae~from the statement and derive the others using
(\ref{jacobi1}). 
\begin{enumerate}
\item
From the relation involving;
$Q_{ja}, Q_{kb}$
and 
$\phi_{l}$ 
we obtain the action of the supercharge
$Q_{ja}$
on
$\psi_{b}$.
\item
The relation involving
$Q_{ja}, Q_{kb}$
and 
$\phi_{l}^{*}$
gives the action of the supercharge
$Q_{ja}$
on
$\chi_{b}$.
\item
The relation involving
$Q_{ja}, \bar{Q}_{k\bar{b}}$
and 
$\phi_{l}$
gives the action of the supercharge
$\bar{Q}_{k\bar{b}}$
on
$\psi_{a}$
and
$\chi_{b}$.
\item
The relation involving
$Q_{ja}, Q_{kb}$
and 
$\bar{\psi}_{\bar{c}}$
gives the action of the central charge $Z$ on
$\psi_{a}$.
\item
The relation involving
$Q_{ja}, Q_{kb}$
and 
$\psi_{c}$
gives the action of the supercharge
$Q_{ja}$
on
$f_{k}$.
\item
The relation involving
$Q_{ja}, \bar{Q}_{k\bar{b}}$
and 
$\psi_{c}$
gives the action of the supercharge
$\bar{Q}_{j\bar{a}}$
on
$f_{k}$.
\item
The relation involving
$Q_{ja}, Q_{kb}$
and 
$\bar{\chi}_{\bar{c}}$
gives the action of the central charge $Z$ on
$\chi_{a}$
\item
The relation involving
$Q_{ja}, Q_{kb}$
and 
$\chi_{c}$
is an identity.
\item
The relation involving
$Q_{ja}, \bar{Q}_{k\bar{b}}$
and 
$\chi_{c}$
is an identity.
\item
The relation involving
$Q_{ja}, Q_{kb}$
and 
$f_{l}$
(or
$f^{*}_{l}$)
gives the action of the central charge $Z$ on
$f_{l}$.
\item
The relation involving
$Q_{ja}, \bar{Q}_{j\bar{b}}$
and 
$f_{c}$
is an identity.
\item
The relations involving a supercharge, the central charge and a fields
are identities. 
\end{enumerate}
Finally one notices the compatibility with the $SU(2)$ transformations of
all the relations from the statement.
$\qed$

Next, we analyze the possible causal (anti)commutation relation. We have
\begin{thm}
Let us suppose that for the multiplet of the preceding Theorem the fields 
$\phi_{j}$ are complex scalar fields of mass 
$m \geq 0$;
in particular we have
\be
~[ \phi_{j}(x), \phi_{k}(y) ] = 0
\ee
\be
~[ \phi_{j}(x), \phi_{k}^{*}(y) ] = - i~\delta_{jk}~D_{m}(x-y).
\ee

Then we also have:
\be
~[ f_{j}(x), f^{*}_{k}(y) ] = -i~m^{2}\delta_{jk}~D_{m}(x-y)
\ee
\be
~[ \phi_{j}(x), f_{k}(y) ] = \alpha \epsilon_{jk}~D_{m}(x-y)
\ee
\be
~[ \phi_{j}(x), f^{*}_{k}(y) ] =  \beta \delta_{jk}~D_{m}(x-y)
\ee
\be
\left\{ \psi_{a}(x), \psi_{b}(y) \right\} = -2~i~\epsilon_{ab}~D_{m}(x-y)
\ee
\be
\left\{ \psi_{a}(x), \bar{\psi}_{\bar{b}}(y) \right\} =
2~\sigma^{\mu}_{a\bar{b}}~\partial_{\mu}D_{m}(x-y)
\ee
\be
\left\{ \chi_{a}(x), \chi_{b}(y) \right\} = 
2 i~\bar{\alpha}~\epsilon_{ab}~~D_{m}(x-y),
\ee
\be
\left\{ \chi_{a}(x), \bar{\chi}_{\bar{b}}(y) \right\} =
2~\sigma^{\mu}_{a\bar{b}}~\partial_{\mu}D_{m}(x-y)
\ee
\be
\left\{ \psi_{a}(x), \chi_{b}(y) \right\} = -2~i~\beta~\epsilon_{ab}~D_{m}(x-y)
\ee
and all other (anti)commutators are zero. Here
$\alpha \in \C$
and
$\beta \in \R$
are free parameters. 
\end{thm}
{\bf Proof:}
From Lorentz and $SU(2)$ covariance considerations we must have:
\be
~[ \phi_{j}(x), f_{k}(y) ] = \alpha \epsilon_{jk}~D_{m}(x-y), \qquad
[ \phi_{j}(x), f^{*}_{k}(y) ] =  \beta \delta_{jk}~D_{m}(x-y)
\ee

Starting from the hypothesis and the preceding relations one determines by some 
computation all the causal (anti)commutators from the Jacobi identity 
(\ref{jacobi2}). 
$\qed$

Concerning the representability in a Hilbert space we have the following 
result:
\begin{thm}
The following multiplet has a representation in a Hilbert space iff
\be
|\alpha|^{2} + \beta^{2} \leq 2.
\ee
\end{thm}
{\bf Proof:}
For the Fermi sector we proceed as in \cite{GS2} i.e we suppose that the Hilbert
space is generated by the Majorana spinors
$f^{(A)}$
verifying the causal anticommutation relations:
\bea
\left\{ f^{(A)}_{a}(x), f^{(B)}_{b}(y) \right\} 
= i~\delta_{AB}~\epsilon_{ab} m~ D_{m}(x-y),
\nonumber \\
\left\{ f^{(A)}_{a}(x), \bar{f}^{(B)}_{\bar{b}}(y) \right\} =
\delta_{AB}~\sigma^{\mu}_{a\bar{b}}~\partial_{\mu}D_{m}(x-y).
\label{CCR-3}
\eea

So, we must have
\be
\psi = \sum_{A} c_{A}~f^{(A)} \qquad
\chi = \sum_{A} d_{A}~f^{(A)}
\label{comp}
\ee
for some (complex) numbers
$
\vec{c} = \{c_{A}\}, \vec{d} = \{d_{A}\}.
$
Like in \cite{GS2} we find out
\bea
\left\{ \psi_{a}(x), \psi_{b}(y) \right\} 
= i~m~\vec{c}^{2}~\epsilon_{ab}~D_{m}(x-y),
\nonumber \\
\left\{ \psi_{a}(x), \bar{\psi}_{\bar{b}}(y) \right\} =
\vec{c}\cdot\vec{c}^{*} \sigma^{\mu}_{a\bar{b}}~\partial_{\mu}D_{m}(x-y)
\nonumber \\
\left\{ \chi_{a}(x), \chi_{b}(y) \right\} 
= i~m~\vec{d}^{2}~\epsilon_{ab}~D_{m}(x-y),
\nonumber \\
\left\{ \chi_{a}(x), \bar{\chi}_{\bar{b}}(y) \right\} =
\vec{d}\cdot\vec{d}^{*} \sigma^{\mu}_{a\bar{b}}~\partial_{\mu}D_{m}(x-y)
\nonumber \\
\left\{ \psi_{a}(x), \chi_{b}(y) \right\} 
= i~m~\vec{c}\cdot\vec{d}~\epsilon_{ab}~D_{m}(x-y),
\nonumber \\
\left\{ \psi_{a}(x), \bar{\chi}_{\bar{b}}(y) \right\} =
\vec{c}\cdot\vec{d}^{*} \sigma^{\mu}_{a\bar{b}}~\partial_{\mu}D_{m}(x-y).
\eea

By comparison we get
\be
m~\vec{c}^{2} = - 2\alpha, \quad
\vec{c}\cdot\vec{c}^{*} = 2, \quad
m~\vec{d}^{2} = 2\bar{\alpha}, \quad
\vec{d}\cdot\vec{d}^{*} = 2, \quad
m~\vec{c}\cdot\vec{d} = - 2\beta, \quad
\vec{c}\cdot\vec{d}^{*} =0.
\ee

We separate the real and the imaginary part of the vectors
$
\vec{c}, \vec{d}
$
and consider all possible Cauchy-Schwartz inequalities. As a result we get
\be
|\alpha|^{2} \leq m^{2}, \qquad
\beta^{2} + \alpha_{1}^{2} \leq m^{2}.
\ee

For the Bosonic sector we simplify the reasoning by some field redefinitions.
If we consider instead of the fields
$f_{j}$
the new fields
\be
F_{j} \equiv f_{j} + c_{1} \phi_{j} + c_{2} \epsilon_{jk} f_{k}^{*} 
+ c_{3} \epsilon_{jk} \phi_{k}^{*} 
\ee
then one can decouple the fields
$\phi_{j}$
from
$F_{j}$
if one chooses
\be
c_{3} = i~(\beta + \bar{\alpha} c_{2}), \qquad
c_{3} = i~(\alpha - \beta c_{2})
\ee 
i.e. with this choice 
$\phi_{j}$
causally commutes with
$F_{j}$
and
$F_{j}^{*}$.
Now we still have to check the positivity for the new fields
$F_{j}$.
One finds out that
\be
[ F_{j}(x), F_{k}(y) ] = -2~i~\alpha\beta \epsilon_{jk}~D_{m}(x-y)
\ee
\be
[ F_{j}(x), F^{*}_{k}(y) ] = -i~(m^{2} - \beta^{2} - |\alpha|^{2})
\delta_{jk}~D_{m}(x-y).
\ee

We now make a new field transformation
\be
g_{j} \equiv F_{j} + c \epsilon_{jk} F_{k}^{*}
\ee
and by a convenient choice of the constant $c$ we arrive at the standard form:
\be
[ g_{j}(x), g_{k}(y) ] = 0
\ee
\be
[ g_{j}(x), g^{*}_{k}(y) ] = - i d~\delta_{jk} D_{m}(x-y)
\ee
for some constant $d$ which can be computed explicitly.
The positivity condition is
$d > 0$
and give the relation from the statement which is stronger than the relation 
obtained in the Fermi sector.
$\qed$

\newpage
\section{The $N= 4$ multiplet\label{n4}}

We consider a model with 
$SU(4)$ 
invariance and no central charges.
\begin{thm}
Let us consider the multiplet 
$(\phi_{jk}, F_{ab}, \lambda_{ja}), \quad j, k = 1,\dots,4$
where:
\begin{itemize}
\item
$\phi_{jk}$
are complex Bosonic scalar fields antisymmetric in $j$ and $k$ and verifying
\be
V_{U}^{-1} \phi_{jk} V_{U} = U_{jl} U_{km} \phi_{lm}, 
\quad \forall U \in SU(4);
\ee
\item
$F_{ab}$
are complex Bosonic fields which are $SU(4)$ scalars;
\item
$\lambda_{ja}$
are spinor Fermionic fields and verifying
\be
V_{U}^{-1} \lambda_{ja} V_{U} = \overline{U_{jk}} \lambda_{k}, 
\quad \forall U \in SO(4).
\ee
\end{itemize}

The the action of the supercharges is well defined by:
\be
i [ Q_{ja}, \phi_{kl} ] = \epsilon_{jklm} \lambda_{ma}
\ee
\be
i [ \bar{Q}_{j\bar{a}} , \phi_{kl} ] = 
\delta_{jk} \bar{\lambda}_{l\bar{a}} - \delta_{jl} \bar{\lambda}_{k\bar{a}}
\ee
\be
\{ Q_{ja}, \lambda_{kb} \} = \delta_{jk} F_{ab}
\ee
\be
\{ Q_{ja}, \bar{\lambda}_{k\bar{b}} \} = 2 \sigma^{\mu}_{a\bar{b}}
\partial_{\mu}\phi_{jk}
\ee
\be
[ Q_{ja} , F_{bc} ] = 0
\ee
\be
[ \bar{Q}_{j\bar{a}}, F_{bc} ] = - 2i \sigma^{\mu}_{c\bar{a}} \partial_{\mu}
\lambda_{kb}
\ee
iff the following constraints are valid:
\be
\partial^{2} \phi_{jk} = 0
\label{ct1}
\ee
\be
\partial^{2} F_{ab} = 0
\label{ct2}
\ee
\be
\sigma^{\mu}_{a\bar{b}}\partial_{\mu} \bar{\lambda}^{\bar{b}}_{j} = 0,
\label{ct3}
\ee
\be
F_{ab} = F_{ba}
\label{ct4}
\ee
\be
\sigma^{\mu}_{a\bar{b}} \epsilon^{ac} \partial_{\mu}F_{cd} = 0.
\label{ct5}
\ee
\be
\epsilon_{jklm} \phi_{lm} = 2 \phi_{jk}^{*}
\label{ct6}
\ee

\end{thm}
{\bf Proof:}
One starts in a well known way from the first two relations and uses 
(\ref{jacobi1}).
\begin{enumerate}
\item
Consider the relation involving;
$Q_{ja}, Q_{kb}$
and 
$\phi^{*}_{lm}$.
If we define
\be
F_{ab} \equiv \{ Q_{jb}, \lambda_{ja} \}
\ee 
we obtain the action of the supercharge
$Q_{ja}$
on
$\lambda_{kb}$
and the symmetry property (\ref{ct4}).
\item
The relation involving
$Q_{ja}, Q_{kb}$
and 
$\phi_{lm}$
is an identity.
\item
The relation involving
$Q_{ja}, \bar{Q}_{k\bar{b}}$
and 
$\phi^{*}_{lm}$
gives the action of the supercharge
$\bar{Q}_{k\bar{b}}$
on
$\lambda_{ja}$
and the constraint (\ref{ct6}).
\item
The relation involving
$Q_{ja}, Q_{kb}$
and 
$\bar{\lambda}_{l\bar{c}}$
gives the constraint (\ref{ct3}).
\item
The relation involving
$Q_{ja}, Q_{kb}$
and 
$\lambda_{lc}$
gives the action of the supercharge
$Q_{ja}$
on
$F_{bc}$.
\item
The relation involving
$Q_{ja}, \bar{Q}_{k\bar{b}}$
and 
$\lambda_{lc}$
gives the action of 
$\bar{Q}_{k\bar{b}}$
on
$F_{ac}$.
\item
The relation involving
$Q_{ja}, Q_{kb}$
and 
$F_{cd}$
is an identity.
\item
The relation involving
$Q_{ja}, Q_{kb}$
and 
$F^{*}_{\bar{c}\bar{d}}$
gives the constraint (\ref{ct1}).
\item
The relation involving
$Q_{ja}, \bar{Q}_{k\bar{b}}$
and 
$F_{cd}$
gives the constraint (\ref{ct5}).
\end{enumerate}

From (\ref{ct3}) and the definition of
$F_{ab}$
given above we also get (\ref{ct2}). Now we take the (anti)commutators of the 
supercharges with the constraints and obtain no new identities. The 
$SU(4)$
consistency of the relation from the statement is easy to obtain.
$\qed$

Let us remark that the anszatz regarding the action of the supercharges on
$\phi_{jk}$
is quite general. If we can consider only 
$SO(4)$ 
covariance a more general situation of the type
\be
i [ Q_{ja}, \phi_{kl} ] = 
\alpha_{1} (\delta_{jk} \lambda_{la} - \delta_{jl} \lambda_{ka})
+ \beta_{1} \epsilon_{jklm} \lambda_{ma}
\ee
\be
i [ \bar{Q}_{j\bar{a}} , \phi_{kl} ] = 
\alpha_{2} (\delta_{jk} \bar{\lambda}_{l\bar{a}} 
- \delta_{jl} \bar{\lambda}_{ka})
+ \beta_{2} \epsilon_{jklm} \bar{\lambda}_{m\bar{a}}
\ee
is possible but one can prove that by clever redefinitions of the fields we 
we can make
$\alpha_{1} = 0 = \beta_{2}$.

To verify the positivity condition it is convenient to replace 
$F_{ab}$
by
$F_{\mu\nu}$
as in Section \ref{n2}; the action of the supercharges on
$F_{\mu\nu}$
is
\be
~[ Q_{ja}, F^{\mu\nu} ] = {i \over 4} 
(\sigma^{\mu}_{a\bar{b}} \partial^{\nu}\bar{\lambda}_{k}^{\bar{b}}
- \sigma^{\nu}_{a\bar{b}} \partial^{\mu}\bar{\lambda}_{k}^{\bar{b}}
- i \epsilon^{\mu\nu\rho\alpha} \sigma_{\alpha a\bar{b}} 
\partial_{\rho}\bar{\lambda}_{j}^{\bar{b}})
\ee
and the field
$F_{\mu\nu}$
must verify the consistency conditions
\be
F_{\mu\nu} = - F_{\nu\mu} \qquad
\partial^{\mu}F_{\mu\nu} = 0 \qquad
\epsilon^{\mu\nu\alpha\beta}\partial_{\beta}F_{\mu\nu} = 0.
\label{ct7}
\ee

Next, we consider the causal (anti)commutator relations. We have:
\begin{thm}
Let us suppose that for the preceding multiplet the field 
$\phi_{jk}$ 
are complex scalar fields of zero mass; in particular we have
\be
~[ \phi_{jk}(x), \phi_{lm}(y) ] = - i 
(\delta_{jl} \delta_{km} - \delta_{jm} \delta_{kl} ) D_{0}(x-y)
\label{phi}
\ee
which is compatible with the 
$SO(4)$
covariance properties. Then we also have:
\be
~[ F_{ab}(x), F_{\bar{c}\bar{d}}^{*}(y) ] = 
2 i~\left( \sigma^{\mu}_{a\bar{c}} \sigma^{\nu}_{b\bar{d}}
+ \sigma^{\mu}_{a\bar{d}} \sigma^{\nu}_{b\bar{c}}\right)~
\partial_{\mu}\partial_{\nu}D_{0}(x-y)
\label{ccr-f}
\ee
\be
\left\{ \lambda_{ja}(x), \bar{\lambda}_{k\bar{b}}(y) \right\} =
2~\delta_{jk}~\sigma^{\mu}_{a\bar{b}}~\partial_{\mu}D_{0}(x-y)
\ee
and all other (anti)commutators are zero. Alternatively, if we work with the 
field
$F_{\mu\nu}$
the relation (\ref{ccr-f}) can be replaced by:
\be
[ F_{\mu\nu}(x), F_{\rho\sigma}(y) ] = - {i\over 2}
( g_{\mu\rho} \partial_{\nu}\partial_{\sigma}
- g_{\nu\rho} \partial_{\mu}\partial_{\sigma}
+ g_{\nu\sigma} \partial_{\mu}\partial_{\rho}
- g_{\nu\rho} \partial_{\mu}\partial_{\sigma})~D_{0}(x-y)
\ee
which is compatible with the constraints (\ref{ct7}). The positivity 
condition is verified for this model.
\end{thm}
{\bf Proof:}
The compatibility of (\ref{phi}) with the 
$SO(4)$
covariance properties is elementary. From Lorentz covariance considerations we 
must have:
\be
~[ \phi_{jk}(x), F_{ab}(y) ] = 0, \qquad
~[ \phi_{jk}^{*}(x), F_{ab}(y) ] = 0
\ee
because there is no partial differential operator
$P_{ab}(\partial)$
symmetric in $a$ and $b$ and Lorentz covariant. 

Starting from the hypothesis and the preceding relations one determines by some 
computation all other causal (anti)commutators from the using 
Jacobi identity (\ref{jacobi2}). One can easily check that these 
(anti)commutation relations are compatible with the constraints (\ref{ct1}) -
(\ref{ct6}). Only the positivity of the field
$F_{\mu\nu}$
requires some work and it is done as in Corollary \ref{pos1}.
$\qed$

We investigate if it is possible to introduce in a consistent way the
the electromagnetic potential
$A_{\mu}$.
The standard construction from the literature - see for instance \cite{So} Sect. 
13, formul\ae~(13.11) - consists of replacing 
$F_{ab}$
by
$A_{\mu}$.
and postulation the following action of the supercharges 
\be
i [ Q_{ja}, \phi_{kl} ] = \delta_{jk} \lambda_{la} - \delta_{jl} \lambda_{ka}
\ee
\be
i [ \bar{Q}_{j\bar{a}} , \phi_{kl} ] = \epsilon_{jklm} \bar{\lambda}_{m\bar{a}}
\ee
\be
\{ Q_{ja}, \lambda_{kb} \} = -i \delta_{jk} \sigma^{\mu\nu}_{ab} F_{\mu\nu}
\ee
\be
\{ Q_{ja}, \bar{\lambda}_{k\bar{b}} \} = 2 \sigma^{\mu}_{a\bar{b}}
\partial_{\mu}\phi_{jk}^{*}
\ee
\be
i [ Q_{ja}, A^{\mu} ] = \sigma^{\mu}_{a\bar{b}} \bar{\lambda}_{j}^{\bar{b}}
\ee
where
\be
F_{\mu\nu} \equiv \partial_{\mu}A_{\nu} - \partial_{\nu}A_{\mu}.  
\ee

One can prove as in \cite{GS2}, Section 8 that SUSY algebra is full-filed on the
physical space
${\cal H}_{phys} = Ker(Q)/Im(Q)$.
However the model does not have a representation in a Hilbert space! The reason
is that the Jacobi identity
\be
[ A^{\mu}(x_{1}), \{\lambda_{ja}(x_{2}), Q_{kb} \} ]
+ \{\lambda_{ja}(x_{2}), [Q_{kb}, A^{\mu}(x_{1}) ] \} = 0
\ee
cannot be satisfied.

Alternatively if we want to solve the constraints (\ref{ct7}) in terms of some 
electromagnetic potential
$A^{\mu}$
we first have from the third constraint that
$
F_{\mu\nu} \equiv \partial_{\mu}A_{\nu} - \partial_{\nu}A_{\mu}
$
and the second constraint gives
$
\partial^{\mu}A_{\mu} = 0;
$
but such a fields does not have a Hilbert space representation.
So it seems that the 
$N = 4$
model considered above cannot be used for a supersymmetric extension of the 
standard model.
\newpage

\section{Conclusions}

The analysis from this paper shows that a quantum supersymmetric
extension of  the standard model using extended supersymmetry is
problematic. The main  reason comes from the severe restrictions on the
causal (anti)commutator  relations  imposed by the positivity condition
(which give a well-defined scalar product  of the model). This is a
serious problem for string theory which seems to predict that the
standard model should be necessarily be an extended supersymmetric one.
For other models the situation is even more dramatic.  Consider for
instance the super-gravity multiplet \cite{Wes} Ch. 9. There one  tries
to extend the linearized Einstein gravity i.e. one describes gravitation
using the field
$h_{\mu\nu}$
symmetric in the indices and which is the first order approximation of the 
metric tensor
$g_{\mu\nu}$.
From the analysis of the irreducible representations of the supersymmetric 
algebra one knows that the multiplet should also contain a Rarita-Schwinger 
field 
$\psi_{a}^{\mu}$
of spin 
$3/2$. 
The theory is considered at the classical level. It particular this means that
in the relations (\ref{tensor}) one should replace the (anti)commutators by an 
action of the supercharges on the supersymmetric manifold with coordinates
$h_{\mu\nu}, \psi_{a}^{\mu}$
and their derivatives. At the level of a classical field theory the 
supersymmetric algebra closes (up to gauge transformations) only if one uses 
the (linearized) Einstein equations. However, if one tries to construct the
corresponding quantum multiplet one finds out that it is not possible to
quantize the field
$h_{\mu\nu}$
such that the Einstein field equations are verified by the quantum operators:
the most general form of the causal commutation relations for
$h_{\mu\nu}$
is not compatible with the (linearized) Einstein equations. The argument 
remains the same even if one introduces the auxiliary fields
$M, N, b_{\mu}$.
(A good quantization procedure for the field
$h_{\mu\nu}$
can be found in \cite{Gr3}, \cite{Sc}). This spoils completely the 
verification of the supersymmetric algebra! 

So, we point out that a quantum construction of supersymmetric multiplet is
more restrictive than the corresponding construction for the classical model.
However, only the quantum model is relevant for any attempt of generalizing
the standard model of elementary particles.

One can see that the restrictions leading to the negative results obtained
in this paper and in the preceding one \cite{GS2} come from the identity
(\ref{jacobi2}). One way to circumvent this restriction is to accept that
the supersymmetry is spontaneously broken. This is in agreement with 
phenomenology which failed to find out supersymmetric partners of the known
elementary particles of equal mass. It is obvious that in a broken 
supersymmetric theory the supercharges 
$Q_{a}, \bar{Q}_{\bar{a}}$
cannot exist. Indeed, the existence of the supercharges and the postulated
relations (\ref{tensor}) lead in all known cases to the equality of the masses
of the Bosons and Fermions. This cannot be saved even if one modifies 
(\ref{tensor}) by adding some constants in the right hand side as it is 
suggested in the literature. The standard literature on spontaneous broken
symmetries suggests indeed that in such a case the charges do not exists but
one hopes to to have the currents as well defined objects and the symmetry 
group of the model is replaced by a current algebra. If such a framework
could be constructed in a supersymmetric model we would expect that 
(\ref{tensor}) are replaced by other relations expressing the (anti)commutation
relations of the supercurrents and the fields. If in (\ref{jacobi1}) and
(\ref{jacobi2}) one replaces the supercharges by the the supercurrents then
less severe restrictions would appear; in particular one would not be
forced to have equal masses. This approach seems worthwhile investigating.
 


\begin{thebibliography}{99}

\bibitem{Bi}
A. Bilal,
``{\it Introduction to Supersymmetry}",
Lecture notes ``Gif 2000", hep-th/0101055

\bibitem{BS}
P. Breitenlohner, M. F. Sohnius,
``{\it An Almost Simple Off-Shell Version of $SU(2)$ Poincar\'e Supergravity}",
Nucl. Phys. {\bf B 178} (1981) 151-176

\bibitem{BSS}
L. Brink, J. H. Schwartz, J. Scherk,
``{\it Supersymmetric Yang-Mills Theories}",
Nucl. Phys. {\bf B 121} 77-92

\bibitem{CS}
F. Constantinescu, G. Scharf,
``{\it Causal Approach to Supersymmetry: Chiral Superfields}",
hep-th/0106090

\bibitem{CGS}
F. Constantinescu, M. Gut, G. Scharf,
``{\it Quantized Superfields}",
Ann. Phys. (Leipzig)
 {\bf 11} (2002) 335-356

\bibitem{IAS}
P. Deligne et. all,
``{\it Quantum Fields and Strings: A Course for Mathematicians}",
vol. 1 \& vol. 2, AMS publ. 2000

\bibitem{Fa}
P. Fayet,
``{\it Fermi-Bose Hypersymmetry}",
Nucl. Phys. {\bf B 113} (1976) 135-155

\bibitem{Gr2} D. R. Grigore,
``{\it Wess-Zumino Model in the Causal Approach}",
hep-th/0011174, European Phys. Journ. {\bf C 21} (2001) 732-734

\bibitem{Gr3} D. R. Grigore,
``{\it On the Quantization of the Linearized Gravitational Field}", \\
hep-th/9905190, Class. Quant. Grav. {\bf 17} (2000) 319-344

\bibitem{GSO}
F. Gliozzi, J. Scherk, D. Olive,
``{\it Supersymmetry, Supergravity Theories and the Dual Spinor Model}",
Nucl. Phys. {\bf B 122} (1977) 253-290 

\bibitem{GGRS}
S. J. Gates Jr., M. T. Grisaru, M.. Ro\u cek, W. Siegel,
``{\it Superspace or One Thousand and One Lessons in Supersymmetry}",
Cummings, 1983, hep-th/0108200 (web edition)

\bibitem{GS1} D. R. Grigore, G. Scharf,
``{\it A Supersymmetric Extension of Quantum Gauge Theory}",
hep-th/0204105, to appear in Annalen der Physik (Leipzig)

\bibitem{GS2} D. R. Grigore, G. Scharf,
``{\it The Quantum Supersymmetric Vector Multiplet
and Some Problems in Non-Abelian Supergauge Theory}",
hep-th/0212026

\bibitem{HST} P. S. Howe, K. S. Stelle, P. K. Townsend,
``{\it The Relaxed Hypermultiplet: an unconstrained $N = 2$ Superfield 
Theory}", 
Nucl. Phys. {\bf B 214} (1983) 519-531

\bibitem{Lo}
J. T. Lopusza\'nski,
``{\it An Introduction to Symmetry and Supersymmetry in Quantum Field Theory}",
World Scientific, 1991

\bibitem{Li}
J. D. Lykken,
``{\it Introduction to Supersymmetry}",
FERMILAB-PUB-96/445-T, \\ hep-th/9612114

\bibitem{LW}
J. T. Lopusza\'nski, M. Wolf,
``{\it Central Charges in the Massive Supersymmetric Quantum Theory of
Scalar-Spinor and Scalar-Spinor-Vector Fields}",
Nucl. Phys. {\bf B 184} (1981) 133-179

\bibitem{NK}
S. Nagamaki, Y. Kobayashi,
``{\it Axioms of Supersymmetric Quantum Fields}", preprint

\bibitem{O}
K. Osterwalder,
``{\it Supersymmetric Quantum Field Theory}" in:
V.Rivasseau (ed.), ``Results in field theory, statistical mechanics and
condensed matter physics", Springer, 
Lectures Notes in Physics {\bf 446} (1995) 117-130

\bibitem{Sc}
G. Scharf,
``{\it Quantum Gauge Theories: A True Ghost Story}",
Wiley, 2001

\bibitem{Sr}
P. P. Srivastava,
``{\it Supersymmetry, Superfields and Supergravity: an Introduction}",
IOP Publ., 1986

\bibitem{So}
M. Sohnius,
``{\it Introducing Supersymmetry}",
Phys. Rep. {\bf 128} (1985) 39-204

\bibitem{SSW1}
M. F. Sohnius, K. S. Stelle, P. C. West,
``{\it Off-Shell Formulation of Extended Supersymmetric Gauge Theories}"
Phys. Lett. {\bf 92 B} (1980) 123-127

\bibitem{SSW2}
M. F. Sohnius, K. S. Stelle, P. C. West,
``{\it Dimensional Reduction by Legendre Transformation Generates Off-Shell
Supersymmetric Yang-Mills Theories}",
Nucl. Phys. {\bf B 173} (1980) 127-153

\bibitem{Wei}
S. Weinberg,
``{\it The Theory of Quantum Fields, vol. 3, Supersymmetry}",
Cambridge Univ. Press, 2000

\bibitem{Wes}
P. West,
``{\it Introduction to Supersymmetry and Supergravity}",
(Extended Second Edition) World Scientific, 1990

\end{thebibliography}
\end{document}